\documentclass{siamart0516}
\usepackage{euscript,amsmath,amssymb,amsfonts,graphicx,bm,mathtools}
  \usepackage{hyperref}
\usepackage{latexsym,amssymb,enumerate,amsmath} 
  \usepackage{graphicx} 
  \usepackage{tikz}
  \usepackage[square,sort,comma,numbers]{natbib}
\usepackage{fancyvrb}

\usepackage{bbm}

\numberwithin{equation}{section}

\newcommand{\dd}{\textup{d}}
\def\eps{\varepsilon}
\def\E{\mathbb{E}}
\def\P{\mathbb{P}}
\def\R{\mathbb{R}}

\newcommand{\x}{\mathbf{x}}

\newcommand{\X}{\mathbf{X}}

\usepackage{amsfonts}
\usepackage{graphicx}
\usepackage{epstopdf}
\usepackage{algorithmic}
\ifpdf
  \DeclareGraphicsExtensions{.eps,.pdf,.png,.jpg}
\else
  \DeclareGraphicsExtensions{.eps}
\fi

\newcommand{\TheTitle}{Yet another look at narrow escape through a tube}
\newcommand{\ShortTitle}{Narrow escape through a tube}
\newcommand{\TheAuthors}{Victorya Richardson, Yick Hin Ling, and Sean D. Lawley}

\headers{\ShortTitle}{\TheAuthors}

\title{{\TheTitle}\thanks{
\funding{The first and third authors were supported by the National Science Foundation (DMS-2325258 and CAREER DMS-1944574). The second author was supported by the Croucher Foundation.}}}

\author{Victorya Richardson\thanks{Department of Mathematics, University of Utah, Salt Lake City, UT 84112 USA.}
\and Yick Hin Ling\thanks{Department of Biology, Johns Hopkins University, Baltimore, MD 21218 USA}
\and Sean D. Lawley\thanks{Department of Mathematics, University of Utah, Salt Lake City, UT 84112 USA (\texttt{lawley@math.utah.edu}).}
}
\date{\today}

\ifpdf
\hypersetup{
  pdftitle={\TheTitle},
  pdfauthor={\TheAuthors}
}
\fi

\begin{document}

\maketitle

\begin{abstract}
The narrow escape problem concerns the time needed for a diffusing particle to exit a confining domain through a small hole in the boundary. While this problem is now well-understood, determining the escape time for a particle that must exit through a narrow tube has proven challenging. Indeed, relying on analogies with electrodynamics, parameter fits to simulations, and heuristics, a variety of conflicting estimates for this escape time have been offered over the last three decades, some of which are counterintuitive and arguably non-physical. In this paper, we combine matched asymptotic analysis and probabilistic methods to determine the exact asymptotics of the narrow escape time through a tube. We obtain a new escape time formula which reduces to the various prior estimates in certain special cases. If the diffusivity in the tube differs from the diffusivity in the rest of the domain, our results reveal the importance of the form of the multiplicative noise inherent to any diffusivity that varies in space. 
We discuss our results in the context of asymmetric cell division.
\end{abstract}



\section{Introduction}

The narrow escape problem is to calculate the average time it takes a diffusing particle to escape from a bounded domain through a small hole in an otherwise reflecting boundary. Owing to (i) its many applications in physical, chemical, and biological problems, and (ii) the rich mathematics entailed in its analysis, the narrow escape problem has attracted great attention from applied mathematicians and physicists in the past two decades (see \cite{grebenkov2019full, cheviakov2010asymptotic, kaye2020, chen2011, lindsay2015, gomez2015, holcman2014, ward10, benichou2008narrow, PB3, lawley2019dtmfpt, isaacson2016uniform} as a sample). For diffusion in three dimensions, this average escape time is \cite{cheviakov2010asymptotic, chen2011},
\begin{align}\label{eq:T0}
    \E[\tau]
    \sim\frac{|\Omega_0|}{4 D a}\quad\text{as }a/|\Omega_0|^{1/3}\to0^+,
\end{align}
where $|\Omega_0|$ is the volume of the confining domain, $D>0$ is the particle diffusivity, and the hole is a disk of radius $a\ll|\Omega_0|^{1/3}$. 

\begin{figure}
    \centering
    \includegraphics[width=0.6\linewidth]{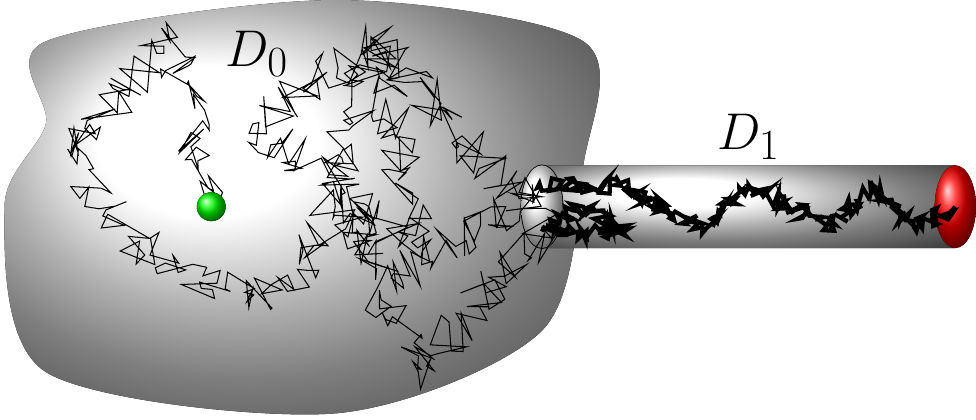}
    \caption{Diffusive path of a particle in a three-dimensional domain consisting of a ``bulk'' volume connected to a tube. The particle starts at the green ball and is eventually absorbed at the red region at the end of the tube. The diffusivity is $D_0$ in the bulk and $D_1$ in the tube}
    \label{fig:schem}
\end{figure}

A series of works have sought to determine the escape time when the diffusing particle must pass through a tube to escape (see Figure~\ref{fig:schem} for an illustration). Much of this work has been in neuroscience \cite{svoboda1996direct, majewska2000regulation, bloodgood2005neuronal, biess2007diffusion}, as researchers have sought to understand how the diffusion of calcium ions and other molecules is regulated by dendritic spine geometry (modeled by a large head connected to the dendritic shaft by a narrow cylindrical neck). A mathematically related problem is to understand the time it takes diffusing molecules to reach ``buried sites'' \cite{samson1978diffusion, zhou1998theory, zhou2010diffusion, berezhkovskii2011diffusion}, which are modeled as absorbing traps located at the end of narrow tubes. In addition, though narrow escape models typically assume that the particle has ``escaped'' as soon as it enters the hole in the boundary, the escape time through a tube is relevant if the particle must pass all the way through the hole to fully escape (i.e.\ if the boundary is not assumed to be infinitesimally thin).

Escape through a tube is also highly relevant to organisms undergoing closed mitosis. Budding yeast asymmetric segregation provides a clear example. During anaphase, the nucleus elongates into a dumbbell shape where two nuclear lobes are connected by a thin tube, which is called the nuclear bridge \cite{boettcher2012nuclear}. This nuclear bridge maintains asymmetric content between the mother and daughter compartments by limiting molecular exchange, which is important for parental identity and asymmetric aging \cite{shcheprova2008mechanism}. Notably, genetic mutants that alter bridge dimensions disrupt this asymmetry \cite{boettcher2012nuclear}. Determining the diffusive escape time through a tube is thus essential for understanding how geometry controls the balance between diffusion and compartmentalization in biological systems. 
Indeed, it is not well understood how compartmentalization in asymmetric cell division depends on (i) pure geometry versus (ii) hindered diffusion in the nuclear bridge (i.e.\ a slower diffusivity in the nuclear bridge compared to the nuclear lobes) \cite{boettcher2012nuclear, zavala2014long}. 

Despite its apparent simplicity, determining the diffusive escape time through a tube has proven challenging. Indeed, a variety of different mathematical approximations have been offered over the past three decades. For a cylindrical tube of length $L$ and radius $a$ that connects to a ``bulk'' domain of volume $|\Omega_0|$, the following escape time was suggested in \cite{svoboda1996direct} using an analogy with electrical currents,
\begin{align}\label{eq:T1}
    \E[\tau]
    \approx\frac{|\Omega_0|L}{\pi a^2 D},
\end{align}
where $D$ is the diffusivity. The formula \eqref{eq:T1} was used in \cite{svoboda1996direct} and later in \cite{majewska2000regulation} and \cite{bloodgood2005neuronal} to interpret neuronal activity data, but the validity of \eqref{eq:T1} was not investigated. 
For this same geometry, the following escape time estimate was derived mathematically from a three-dimensional diffusion model \cite{holcman2004escape, singer2006narrow, singer2006narrow2, schuss_narrow_2007},
\begin{align}\label{eq:T2}
    \E[\tau]
    \approx\frac{|\Omega_0|}{4 D a}+\frac{L^2}{2D}.
\end{align}
The estimate \eqref{eq:T2} assumes that the particle cannot return to the bulk $\Omega_0$ after entering the tube, and thus the escape time in \eqref{eq:T2} is the time to exit the bulk (i.e.\ $|\Omega_0|/(4 D a)$ in \eqref{eq:T0}) plus the time to escape a one-dimensional tube of length $L$ (i.e.\ $L^2/(2D)$). Relaxing this simplifying `no return' assumption, the following estimate was later posited in \cite{biess2007diffusion} by summing over the number of times that the particle returns to the bulk,
\begin{align}\label{eq:T3}
    \E[\tau]
    \approx \frac{1}{\beta}\frac{|\Omega_0|L}{4Da^2}
    +\frac{L^2}{2D},
\end{align}
where $\beta=0.84$ was obtained by fitting the functional form \eqref{eq:T3} to stochastic simulations. Note that $4\beta=3.36\approx1.07\pi$, and thus the first term in \eqref{eq:T3} differs from \eqref{eq:T1} by around 7\%. 
It was later shown in \cite{berezhkovskii2009escape} that the sum of \eqref{eq:T1} and \eqref{eq:T2} agreed with some numerical estimates of $\E[\tau]$ obtained by numerically solving the diffusion equation describing the escape time with a finite element method. In fact, Ref.~\cite{berezhkovskii2009escape} generalized the sum of \eqref{eq:T1} and \eqref{eq:T2} to allow the particle to have different diffusivities in the bulk and the tube,
\begin{align}\label{eq:T4}
    \E[\tau]
    \approx\frac{|\Omega_0|L}{\pi a^2 D_1}
    +\frac{|\Omega|}{4 D_0 a}+\frac{L^2}{2D_1},
\end{align}
where $D_0$ is the diffusivity in the bulk and $D_1$ is the diffusivity in the tube. The formula in \eqref{eq:T4} was derived by positing an elegant formalism in which (i) particles enter the tube from the bulk at exponentially distributed times with mean given by \eqref{eq:T0}, (ii) particles move in the tube according to a one-dimensional diffusion equation, and (iii) particles can return to the bulk from the tube via a partially absorbing boundary condition at their interface.

Reviewing the prior estimates in \eqref{eq:T1}-\eqref{eq:T4} raises several questions. Can the narrow escape time through a tube be determined more precisely? For narrow escape through a hole (rather than a tube), the estimate in \eqref{eq:T0} can be obtained by studying the governing three-dimensional diffusion equation via the method of matched asymptotic expansions \cite{cheviakov2010asymptotic} or rigorous analytical methods \cite{chen2011} (and higher order corrections to \eqref{eq:T0} can be obtained with similar systematic methods). In contrast, \eqref{eq:T1}-\eqref{eq:T4} rely on electrical analogies, heuristics, and/or parameter fits to numerical simulations.

Furthermore, can a single formula be derived which remains valid in parameter regimes where \eqref{eq:T1}-\eqref{eq:T4} break down? For instance, the times in \eqref{eq:T1} and \eqref{eq:T3} both vanish if the tube length $L$ vanishes, but the escape time should reduce to the ``narrow hole'' time $|\Omega_0|/(4Da)$ in \eqref{eq:T0} in this limit. While \eqref{eq:T2} has the (presumably) correct asymptotic behavior when $L$ vanishes, \eqref{eq:T2} diverges like $1/a$ if the tube radius $a$ vanishes, whereas \eqref{eq:T1}, \eqref{eq:T3}, and \eqref{eq:T4} diverge like $1/a^2$. 
Furthermore, the predictions of \eqref{eq:T4} are counterintuitive if the bulk and tube diffusivities are not equal (i.e.\ if $D_0\neq D_1$). For instance, \eqref{eq:T4} predicts that the bulk diffusivity $D_0$ is irrelevant if the tube radius $a$ vanishes. In addition, one might expect the escape time to reduce to $L^2/D_1$ if $D_0\to\infty$, but this contradicts \eqref{eq:T4}. Furthermore, suppose the timescale of diffusion in the tube vanishes, $L^2/D_1\to0$. It is natural to expect the escape time to reduce to the narrow hole time $|\Omega_0|/(4Da)$ in \eqref{eq:T0} in this limit, but the time in \eqref{eq:T4} could diverge in this limit (for instance, if $L\propto\eps$ and $D_1\propto\eps^{3/2}$, then $L^2/D_1\to0$ but $L/D_1\to\infty$ for $\eps\to0$, and thus the time in \eqref{eq:T4} diverges). Can these predictions of \eqref{eq:T4} be correct?

The purpose of this paper is to answer these questions. By combining matched asymptotic analysis \cite{cheviakov2010asymptotic, ward93} with probabilistic techniques, we determine the exact asymptotics of $\E[\tau]$ as the tube radius $a$ vanishes compared to the lengthscales of the bulk $\Omega_0$ (i.e.\ as $a/|\Omega_0|^{1/3}\to0$ as in \eqref{eq:T0}) under various scalings of the tube length $L$, bulk diffusivity $D_0$, and tube diffusivity $D_1$. Furthermore, we suggest a single formula that accurately approximates $\E[\tau]$ for vanishing tube radius $a$ and arbitrary values of $L$, $D_0$, and $D_1$.

If $D_0\neq D_1$, then our results depend crucially on the parameter ${\alpha}\in[0,1]$ that specifies the form of the multiplicative noise inherent to any diffusivity that depends on space. That is, if the diffusivity changes from $D_0$ in the bulk to $D_1$ in the tube (with $D_0\neq D_1$), then one must necessarily specify $\alpha\in[0,1]$. The most common choices in the literature are ${\alpha}=0$ (It\^{o} calculus \cite{ito1944stochastic}) and ${\alpha}=1/2$ (Stratonovich calculus \cite{stratonovich1966new}) \cite{mannella2012ito}. The so-called ``It\^{o} versus Stratonovich'' controversy sparked debate for almost a decade, but it was eventually settled that there is no universally ``correct'' choice of ${\alpha}$ \cite{van1981ito, mannella2012ito}. That is, the parameter ${\alpha}\in[0,1]$ is part of the model and must be chosen on physical grounds \cite{van1981ito, mannella2012ito}.

We now summarize our results. To compare with \eqref{eq:T1}-\eqref{eq:T4}, suppose that the tube is a cylinder with radius $a$ and length $L$ (though we establish our results for tubes of general cross-sectional shape). Assume that the particle starts in the bulk away from the entrance to the tube. Decompose the mean escape time into the mean residence time in the bulk, $\E[\tau_0]$, plus the mean residence time in the tube, $\E[\tau_1]$,
\begin{align*}
    \E[\tau]
    =\E[\tau_0]+\E[\tau_1].
\end{align*}
We first show the exact equality $\E[\tau_1]=L^2/(2D_1)$. We then derive the following exact asymptotic escape time,
\begin{align}\label{eq:mfpt0}
    \E[\tau]
    \sim\frac{1}{C(L/a,(D_0/D_1)^{\alpha})}\frac{|\Omega_0|}{2\pi D_0a}
    +\frac{L^2}{2D_1}\quad\text{as }a/|\Omega_0|^{1/3}\to0,
\end{align}
where $C=C(L/a,(D_0/D_1)^{\alpha})$ is determined by a certain ``inner problem'' that arises in the matched asymptotic analysis. In words, $C(L/a,(D_0/D_1)^{\alpha})$ is the ``capacitance'' of a unit disk that is in a ``pit'' of depth $L/a$ below a reflecting surface, where the diffusivity is $D_0$ above the pit and $D_1$ in the pit, and the change in diffusivity is interpreted with parameter ${\alpha}\in[0,1]$. We further obtain the following exact asymptotics of $C$,
\begin{alignat}{2}
    C(L/a,(D_0/D_1)^{\alpha})
    &\to2/\pi\quad&&\text{as }\rho\to0,\label{eq:Casymptotic0}\\
    C(L/a,(D_0/D_1)^{\alpha})
    &\sim1/(2\rho)
    \quad&&\text{as }\rho\to\infty,\label{eq:Casymptotic1}
\end{alignat}
where the dimensionless parameter
\begin{align*}
    \rho
    :=(L/a)(D_0/D_1)^{\alpha}>0
\end{align*}
combines the tube aspect ratio $L/a$ and the change in diffusivity $(D_0/D_1)^{\alpha}$. If $\alpha\neq0$, notice that we can have $\rho\ll1$ if the tube is short ($L/a\ll1$) and/or has fast diffusivity ($(D_0/D_1)^\alpha\ll1$). Similarly, if $\alpha\neq0$, then we can have $\rho\gg1$ if the tube is long and/or has slow diffusivity. Combining \eqref{eq:mfpt0} with \eqref{eq:Casymptotic0}-\eqref{eq:Casymptotic1} yields the mean escape time estimates
\begin{align*}
    \E[\tau]
    &\approx\frac{|\Omega_0|}{4 D_0a}
    +\frac{L^2}{2D_1}\,\quad\qquad\quad\text{if }a/|\Omega_0|^{1/3}\ll1,\,\rho\ll1,\\
    \E[\tau]
    &\approx\frac{|\Omega_0|L}{\pi a^2 D_0^{1-\alpha}D_1^\alpha}
    +\frac{L^2}{2D_1}\quad\text{if }a/|\Omega_0|^{1/3}\ll1,\,\rho\gg1.
\end{align*}
Furthermore, we use kinetic Monte Carlo simulations to show that $C(L/a,(D_0/D_1)^{\alpha})$ is well-approximated by the following sigmoidal function which interpolates between the two limits in \eqref{eq:Casymptotic0}-\eqref{eq:Casymptotic1},
\begin{align}\label{eq:sigmoid0}
    C(L/a,(D_0/D_1)^{\alpha})
    \approx\frac{2/\pi}{1+(4/\pi)\rho},\quad\text{for all }\rho>0.
\end{align}
Combining \eqref{eq:sigmoid0} with \eqref{eq:mfpt0} yields the following mean escape time estimate that is accurate for all $\rho>0$ as $a/|\Omega_0|^{1/3}$ vanishes,
\begin{align*}
    \E[\tau]
    &\approx\frac{|\Omega_0|L}{\pi a^2 D_0^{1-\alpha}D_1^\alpha}
    +\frac{|\Omega_0|}{4 D_0a}
    +\frac{L^2}{2D_1}\quad\text{if }a/|\Omega_0|^{1/3}\ll1.
\end{align*}

The rest of the paper is organized as follows. Section~\ref{sec:core} is devoted to the aforementioned inner problem. 
Section~\ref{sec:core} contains the inner problem's (i) formulation as a partial differential equation (PDE) boundary value problem, (ii) probabilistic representation, (iii) asymptotics, and (iv) comparison to kinetic Monte Carlo simulations. Relying on this inner problem analysis, Section~\ref{sec:mfpt} uses matched asymptotics \cite{cheviakov2010asymptotic, ward93} to study the narrow escape time through a tube. We conclude in section~\ref{sec:discussion} by (a) describing relations to previous work, including the prior estimates in \eqref{eq:T1}-\eqref{eq:T4} and partially absorbing traps \cite{chaigneau2022first, grebenkov2017effects, grebenkov2017diffusive, grebenkov2019full, guerin2023imperfect, grebenkov2025asymptotic}, and (b) discussing our results in the context of asymmetric protein segregation in closed mitosis.

\section{Inner problem}\label{sec:core}

We now present some probabilistic analysis of a certain PDE boundary value problem. This problem arises in our study of the ``inner problem'' in the matched asymptotic analysis in section~\ref{sec:mfpt} below. 

\subsection{PDE}

Fix a depth $\delta\in(0,\infty)$ and a simply connected open set $\Gamma\in\R^2$ with boundary $\partial\Gamma$ and finite area $|\Gamma|\in(0,\infty)$. We are most interested in the case that $\Gamma$ is a disk, but we can perform most of our analysis for more general shapes. Suppose that $u=u(x,y,z)$ satisfies
\begin{align}\label{eq:core}
\begin{split}
    \Delta u
    &=0, \quad \{(x,y,z)\in\R^3:x>0\}\cup\{(x,y,z)\in(-\delta,0)\times\Gamma\},\\
    u
    &=1, \quad x=-\delta,\,(y,z)\in\Gamma,\\
    \partial_x u
    &=0, \quad x=0,\,(y,z)\notin\overline{\Gamma},\\
    \partial_{n}u
    &=0,\quad x\in(-\delta,0),\,(y,z)\in\partial\Gamma,
\end{split}    
\end{align}
where $\Delta=\partial_{xx}+\partial_{yy}+\partial_{zz}$ is the Laplacian and $\partial_n$ is the outward normal derivative to $\partial\Gamma$. In words, the function $u$ is harmonic in the union of upper half space with a ``tube'' or ``pit'' of depth $\delta\ge0$ and cross-sectional shape $\Gamma$. Furthermore, $u$ satisfies a unit Dirichlet boundary condition at the bottom of the pit and reflecting boundary conditions everywhere else. Using $f(y\pm):=\lim_{x\to y\pm}f(x)$ to denote one-sided limits, we further impose the following continuity conditions \cite{vaccario2015, bressloff2017temporal},
\begin{align}\label{eq:cont2}
\begin{split}
    u(0^-,y,z)
    &=u(0^+,y,z),\quad (y,z)\in\Gamma,\\
    D_1^{\alpha}\partial_x u(0^-,y,z)
    &=D_0^{\alpha}\partial_x u(0^+,y,z),\quad (y,z)\in\Gamma,
\end{split}    
\end{align}
for given diffusivities $D_0>0$ and $D_1>0$ and a given multiplicative noise parameter ${\alpha}\in[0,1]$. 

\subsection{Probabilistic interpretation}\label{sec:probinterp}

The solution $u=u(\x)=u(x,y,z)$ of \eqref{eq:core}-\eqref{eq:cont2} is the probability that a diffusing particle that starts at $\x=(x,y,z)$ will eventually hit the bottom of the tube. More precisely, consider a path $\X=\{\X(t)\}_{t\ge0}$ that reflects from the boundary of the domain in \eqref{eq:core} and diffuses according to the following stochastic differential equation written in term of its infinitesimal increments,
\begin{align*}
    \X(t+\dd t)
    = \X(t)+\sqrt{2D(\X_*)}\,\dd \mathbf{W},
\end{align*}
where $\mathbf{W}=\{\mathbf{W}(t)\}_{t\ge0}$ is a standard three-dimensional Brownian motion, $D:\R^3\mapsto(0,\infty)$ is a given space-dependent diffusivity, and  $\X_*$ is the following weighted average of $\X(t)$ and $\X(t+\dd t)$,
\begin{align*}
    \X_*
    =(1-{\alpha})\X(t)+{\alpha} \X(t+\dd t),
\end{align*}
for a given parameter ${\alpha}\in[0,1]$. Importantly, the path $\X$ depends on the choice of ${\alpha}\in[0,1]$ (unless the diffusivity is constant $D(\x)\equiv D$). Common choices are ${\alpha}=0$ (It\^{o}), ${\alpha}=1/2$ (Stratonovich), and ${\alpha}=1$ (isothermal). The probability density
\begin{align*}
    p(x,y,z,t)
    =\frac{\P(\X(t)=(x,y,z))}{\dd x\,\dd y\,\dd z}
\end{align*}
evolves according to the following forward Fokker-Planck equation \cite{vaccario2015, bressloff2017temporal},
\begin{align*}
    \partial_t p
    =\nabla\cdot[D^{\alpha}\nabla\cdot[D^{1-{\alpha} }p]].
\end{align*}
If we take the diffusivity to be $D_0$ outside the tube and $D_1$ inside the tube,
\begin{align*}
    D(x,y,z)
    =\begin{cases}
        D_0 & \text{if }x>0,\\
        D_1 & \text{if }x<0,\,(y,z)\in\Gamma,
    \end{cases}
\end{align*}
then the solution $u=u(x,y,z)=u(\x)$ of \eqref{eq:core} is
\begin{align*}
    u(\x)
    =\P(\tau<\infty\,|\,\X(0)=\x),
\end{align*}
where $\tau:=\inf\{t\ge0:\X(t)\in\{-\delta\}\times\Gamma\}$ is the first time that $\X$ hits the bottom of the tube.

\subsection{Far-field behavior}

We claim that $u$ has the following monopole decay at far-field,
\begin{align}\label{eq:corefar}
    u\sim\frac{C(\Gamma,\delta,(D_0/D_1)^{\alpha})}{\sqrt{x^2+y^2+z^2}}\quad\text{as }\sqrt{x^2+y^2+z^2}\to\infty,
\end{align}
for a function $C=C(\Gamma,\delta,(D_0/D_1)^{\alpha})$ of the depth $\delta$, shape $\Gamma$, and ratio $(D_0/D_1)^{\alpha}$. The monopole $C$ in \eqref{eq:corefar} can be interpreted as the ``electrostatic capacitance'' of the shape $\Gamma$ that is buried in a pit of depth $\delta>0$, where the diffusivity changes from $D_0$ to $D_1$ upon entering the pit, and this change in diffusivity is interpreted via ${\alpha}\in[0,1]$.

To see that \eqref{eq:corefar} holds, note first that the strong Markov property implies that
\begin{align}\label{eq:strongMarkov}
\begin{split}
    \P(\tau<\infty\,|\,\X(0)=\x)
    &=\P(\tau_R<\infty\,|\,\X(0)=\x)\int_{\partial H_R} u(\x')\rho_R(\x'|\x)\,\dd \x',\quad \|\x\|>R,
\end{split}
\end{align}
where $\tau_R:=\inf\{t\ge0:\X(t)\in  H_R\}$ is the first time that $\X$ hits the hemisphere
\begin{align}\label{eq:hemi}
H_R:=\{(x',y',z')\in\R^3:\|(x',y',z')\|< R,\,x'>0\},    
\end{align}
whose radius is anything large enough that it encapsulates the top of the pit (i.e.\ $y^2+z^2<R^2$ for all $(y,z)\in\Gamma$), and $\rho_R(\x'|\x)$ is the distribution of first hitting position to the boundary of the hemisphere $\partial H_R$ for a particle starting from $\x$ (conditioned that the particle hits $H_R$). It is well-known that $\P(\tau_0<\infty\,|\,\X(0)=\x)$ is a harmonic function of $\x$ that vanishes as $\|\x\|\to\infty$ and is equal to unity if $\|\x\|=R$. Hence,
\begin{align*}
    \P(\tau_0<\infty\,|\,\X(0)=\x)
    =\frac{R}{\|\x\|},\quad\text{if }\|\x\|\ge R>0.
\end{align*}
Taking $\|\x\|\to\infty$ in \eqref{eq:strongMarkov} yields \eqref{eq:corefar}. Furthermore, it follows from symmetry that as $\|\x\|\to\infty$, the hitting distribution $\rho_R(\x'|\x)$ becomes uniform on the upper part of the boundary of the hemisphere,
\begin{align*}
\partial H_R^+
:=\{(x',y',z')\in\R^3:\|(x',y',z')\|=R,\,x'\ge0\}.
\end{align*}
Thus, taking $\|\x\|\to\infty$ in \eqref{eq:strongMarkov} implies that \eqref{eq:corefar} holds with monopole given in terms of the following integral of $u$ over the surface $\partial H_R^+$,
\begin{align}\label{eq:rela}
    C
    =C(\Gamma,\delta,(D_0/D_1)^{\alpha})
    =R\Big(\frac{1}{2\pi R^2}\int_{\partial H_R^+}u\,\dd S\Big).
\end{align}
We use the relation \eqref{eq:rela} in section~\ref{sec:kmccore} to estimate $C(\Gamma,\delta,(D_0/D_1)^{\alpha})$ from stochastic simulations of $\X$ that start uniformly distributed on $\partial H_R^+$. Note also that \eqref{eq:rela} gives the following simple bound since $u\le1$,
\begin{align}\label{eq:Cbound}
    C
    =C(\Gamma,\delta,(D_0/D_1)^{\alpha})
    \le R\quad\text{for all }\delta\ge0,\,(D_0/D_1)^{\alpha}>0.
\end{align}

\subsection{Inner problem asymptotics}

We now use probability theory to determine the asymptotics of the monopole $C(\Gamma,\delta,(D_0/D_1)^{\alpha})$. We used a simpler version of this argument in the Appendix of \cite{richardson2025quantifying} for the case that $D_0=D_1$ and $\Gamma$ is a square. 
Consider $\X$ starting uniformly distributed in the lateral directions of $\Gamma$ at a ``height'' $x\in[-\delta,\infty)$,
\begin{align*}
    \overline{u}(x)
    :=\frac{1}{|\Gamma|}\int_{\Gamma}u(x,y,z)\,\dd y\,\dd z.
\end{align*}
If $x\in[-\delta,0)$ (i.e.\ the height is below the surface), then it follows from the strong Markov property and splitting probabilities for a one-dimensional Brownian motion that
\begin{align}\label{eq:split0}
    \overline{u}(x)
    =|x|/\delta+(1-|x|/\delta)\overline{u}(0),\quad \text{if }x\in[-\delta,0).
\end{align}
In words, the first term in the right hand side of \eqref{eq:split0} is the probability that the particle hits the bottom of the pit before hitting the top of the pit. The second term is the complementary probability, $1-|x|/\delta$, multiplied by the probability that the particle eventually hits the bottom of the pit having started at the top of the pit, $\overline{u}(0)$. Importantly, since the particle is initially uniformly distributed in the lateral directions of $\Gamma$, this distribution remains uniform until exiting the pit. Differentiating \eqref{eq:split0} in $x$ yields 
\begin{align}\label{eq:split}
    \partial_x \overline{u}(x)
    =-(1-\overline{u}(0))/\delta<0,\quad\text{if }x\in[-\delta,0).
\end{align}

Let $H_R\in\R^3$ be the hemisphere in \eqref{eq:hemi}.
Integrating $\Delta u$ over $H_R$ and using \eqref{eq:core} and the divergence theorem yields
\begin{align}\label{eq:divergence0}
    0
    =\int_{H_R} \Delta u\,\dd V
    = \int_{\partial H_R^+}\partial_n u\,\dd S
    -\int_{\Gamma}\partial_x u(0^+,y,z)\,\dd y\,\dd z,
\end{align}
where we have used the reflecting boundary conditions in \eqref{eq:core} to eliminate the boundary integrals on the reflecting parts of the boundary. The far-field behavior in \eqref{eq:corefar} implies that as $R\to\infty$, the surface integral over $\partial H_R^+$ in \eqref{eq:divergence0} limits to
\begin{align}\label{eq:outerboundary}
    \int_{\partial H_R^+}\partial_n u\,\dd S
    \sim(2\pi R^2)\Big(\frac{-C}{R^2}\Big)
    =-2\pi C\quad\text{as }R\to\infty.
\end{align}
To determine the second term in \eqref{eq:divergence0}, we use the second continuity condition in \eqref{eq:cont2} to obtain
\begin{align}\label{eq:continuityintegral}
\begin{split}
    \int_{\Gamma}\partial_x u(0^+,y,z)\,\dd y\,\dd z
    &=D_0^{-{\alpha}} D_1^{\alpha} \int_{\Gamma}\partial_x u(0^-,y,z)\,\dd y\,\dd z\\
    &=D_0^{-{\alpha}} D_1^{\alpha}|\Gamma|\partial_x \overline{u}(0^-)
    =-D_0^{-{\alpha}} D_1^{\alpha}|\Gamma|(1-\overline{u}(0))/\delta,
\end{split}    
\end{align}
where we have interchanged differentiation and integration to obtain the second equality and used \eqref{eq:split} to obtain the third equality. 

Combining \eqref{eq:divergence0}, \eqref{eq:outerboundary}, and \eqref{eq:continuityintegral} yields
\begin{align*}
    0
    =-2\pi C
    +D_0^{-{\alpha}}D_1^{\alpha}|\Gamma|(1-\overline{u}(0))/\delta,
\end{align*}
and rearranging yields the following exact representation for the monopole $C$ in \eqref{eq:corefar},
\begin{align}\label{eq:exact}
    C=C(\Gamma,\delta,(D_0/D_1)^{\alpha})
    =\frac{|\Gamma|(1-\overline{u}(0))}{2\pi (D_0/D_1)^{\alpha}}\frac{1}{\delta}.
\end{align}
Though \eqref{eq:exact} is an exact equation that holds for all $\delta>0$, we do not know the value of $\overline{u}(0)$ or its exact dependence on $\delta$, $\Gamma$, and $(D_0/D_1)^{\alpha}$. However, we can use \eqref{eq:exact} to determine the asymptotics of $C=C(\Gamma,\delta,(D_0/D_1)^\alpha)$.

\subsubsection{Taking $\rho:=(D_0/D_1)^{\alpha}\delta\to0$}
First, suppose we take $\rho:=(D_0/D_1)^{\alpha}\delta\to0$ in \eqref{eq:exact}. Since $C$ is bounded by \eqref{eq:Cbound}, the relation \eqref{eq:exact} ensures that
\begin{align*}
    \overline{u}(0)
    =\frac{1}{|\Gamma|}\int_\Gamma u(0,x,y)\,\dd y\,\dd z\to1\quad\text{as }\rho:=(D_0/D_1)^\alpha\delta\to0.
\end{align*}
Since $u\le1$, we thus have pointwise convergence,
\begin{align}\label{eq:to1}
    u(0,y,z)\to1\quad\text{as $\rho\to0$ for all }(y,z)\in\Gamma.
\end{align}
Furthermore, combining the strong Markov property with \eqref{eq:rela} implies that
\begin{align}\label{eq:CC0}
    \frac{C}{R}
    =\frac{1}{2\pi R^2}\int_{\partial H_R^+}u\,\dd S
    =\frac{C_0(\Gamma)}{R}\int_\Gamma u(0,y,z)p_R(y,z)\,\dd y\,\dd z,
\end{align}
where $p_R$ is the probability distribution of where the particle hits $\Gamma$, given that the particle hits $\Gamma$ after starting uniformly distributed on $\partial H_R^+$, and $C_0(\Gamma)=C(\Gamma,0,1)$ is the electrostatic capacitance of $\Gamma$. That is, $C_0(\Gamma)$ is given in terms of the probability that the particle hits $\Gamma$ given that it starts uniformly on a hemisphere containing $\Gamma$,
\begin{align}\label{eq:cap}
    C_0(\Gamma)
    =R\,\P\big(\tau_\Gamma<\infty\,|\,\X(0)=_\dd \textup{uniform}(\partial H_R^+)\big),
\end{align}
where 
$\tau_\Gamma:=\inf\{t\ge0:\X(t)\in\{0\}\times\Gamma\}$ is the first hitting time to $\Gamma$. Taking $\rho\to0$ in \eqref{eq:CC0} and using \eqref{eq:to1} yields the small $\rho$ asymptotics of $C$,
\begin{align}\label{eq:shallow}
    C=C(\Gamma,\delta,(D_0/D_1)^{\alpha})
    \to C_0(\Gamma)\quad\text{as }\rho:=(D_0/D_1)^\alpha\delta\to0.
\end{align}

\subsubsection{Taking $\rho:=(D_0/D_1)^{\alpha}\delta\to\infty$}
To obtain the large $\rho$ asymptotics of $C$, we first claim that 
\begin{align}\label{eq:to0}
         u(0,y,z)\to0\quad\text{as $\rho:=(D_0/D_1)^{\alpha}\delta\to\infty$ for all }(y,z)\in\Gamma.
\end{align}
To prove \eqref{eq:to0}, consider a one-dimensional diffusion process with space-dependent diffusivity $D(x)=D_0>0$ if $x>0$ and $D(x)=D_1>0$ if $x<0$, where the change in diffusivity at $x=0$ is interpreted with multiplicative noise parameter ${\alpha}$. If $P(x_0)=P(x_0;\delta,\eta)$ is the probability that the process reaches $x=-\delta<0$ before $x=\eta>0$ given that it starts at $x_0\in[-\delta,\eta]$, then $P(x_0)$ satisfies
\begin{align*}
    P''=0,\quad\text{if }x\in(-\delta,0)\cup(0,\eta),
\end{align*}
with continuity conditions at $x=0$,
\begin{align*}
    P(0^-)
    &=P(0^+),\\
    D_0^{\alpha} P'(0^-)
    &=D_1^{\alpha} P'(0^+),
\end{align*}
and boundary conditions $P(-\delta)=1$ and $P(\eta)=0$. Hence, 
\begin{align}\label{eq:P}
\begin{split}
    P(x_0)
    &=\begin{cases}
        1-\frac{1 +x/\delta}{1 +\eta/\rho} & \text{if }x\in[-\delta,0],\\
        \frac{1 -x/\eta}{1+\rho/\eta } & \text{if }x\in[0,\eta],
    \end{cases}
\end{split}    
\end{align}
where $\rho=(D_0/D_1)^{\alpha}\delta>0$. 

Let $\tau_0=\inf\{t\ge0:X(t)\in\{-\delta,\eta\}\}$ be the first time that $X$ hits either $x=-\delta<0$ or $x=\eta>0$. The strong Markov property implies that
\begin{align}\label{eq:keybound}
\begin{split}
    u(0,y,z)
    &\le\P(X(\tau_0)=-\delta\,|\,\X(0)=(0,y,z))+\sup_{(y',z')}u(\eta,y',z')\\
    &\le P(0)+\sup_{(y',z')}u(\eta,y',z').
\end{split}    
\end{align}
Now, the decay in \eqref{eq:corefar} ensures that
\begin{align*}
    \lim_{\eta\to\infty}\sup_{(y',z')\in\R^2}u(\eta,y',z')=0.
\end{align*}
Therefore, taking $\eta=\sqrt{\rho}\to\infty$ in \eqref{eq:keybound} and using that $P(0)=1/(1+\rho/\eta)$ from \eqref{eq:P} yields \eqref{eq:to0}.

Taking $\rho:=(D_0/D_1)^{\alpha}\delta\to\infty$ in \eqref{eq:exact} and using \eqref{eq:to0} yields the following asymptotics of $C$,
\begin{align}\label{eq:deep}
    C(\Gamma,\delta,(D_0/D_1)^{\alpha})
    \sim\frac{|\Gamma|}{2\pi (D_0/D_1)^{\alpha}}\frac{1}{\delta}
    =\frac{|\Gamma|}{2\pi\rho}\quad\text{as }\rho:=(D_0/D_1)^{\alpha}\delta\to\infty.
\end{align}
Note that the shape $\Gamma$ affects \eqref{eq:deep} solely via its area $|\Gamma|$. 

\subsection{Sigmoid approximation}

Having determined the small $\rho$ and large $\rho$ asymptotics of $C$ in \eqref{eq:shallow} and \eqref{eq:deep}, we propose the following heuristic sigmoid approximation to $C$ which interpolates between these two limits,
\begin{align}\label{eq:sigmoid}
    C(\Gamma,\delta,(D_0/D_1)^{\alpha})
    \approx C_{\mathrm{sigmoid}}(\Gamma,\rho)
    :=\frac{C_0(\Gamma)}{1+2\pi\rho C_0(\Gamma)/|\Gamma|},\quad \rho:=(D_0/D_1)^{\alpha}\delta\ge0,
\end{align}
where $C_0(\Gamma)=C(\Gamma,0,1)$ is the electrostatic capacitance of $\Gamma$ in \eqref{eq:cap}, which is known for some choices of the shape $\Gamma$. For instance, if $\Gamma$ is a disk of area $|\Gamma|=\pi r^2$, then \cite{jackson1975}
\begin{align*}
    C_0(\Gamma)
    =\frac{2r}{\pi}\quad\text{if }\Gamma=\{(y,z):y^2+z^2<r^2\},
\end{align*}
and thus
\begin{align}\label{eq:sigmoiddisk}
    C_{\mathrm{sigmoid}}(\Gamma,\rho)
    =\frac{(2/\pi)r}{1+(4/\pi)\rho/r}
    \quad\text{if }\Gamma=\{(y,z):y^2+z^2<r^2\},
\end{align}

Though the sigmoid approximation $C_{\mathrm{sigmoid}}$ in \eqref{eq:sigmoid} has the correct large and small $\delta$ behavior, the interpolation between these asymptotics is heuristic. Nevertheless, we show below using stochastic simulations that $C_{\mathrm{sigmoid}}$ accurately approximates $C(\Gamma,\delta,(D_0/D_1)^{\alpha})$ for all $\delta\ge0$ if $\Gamma$ is a disk.

\subsection{KMC simulations}\label{sec:kmccore}

\begin{figure}
    \centering
    \includegraphics[width=0.6\linewidth]{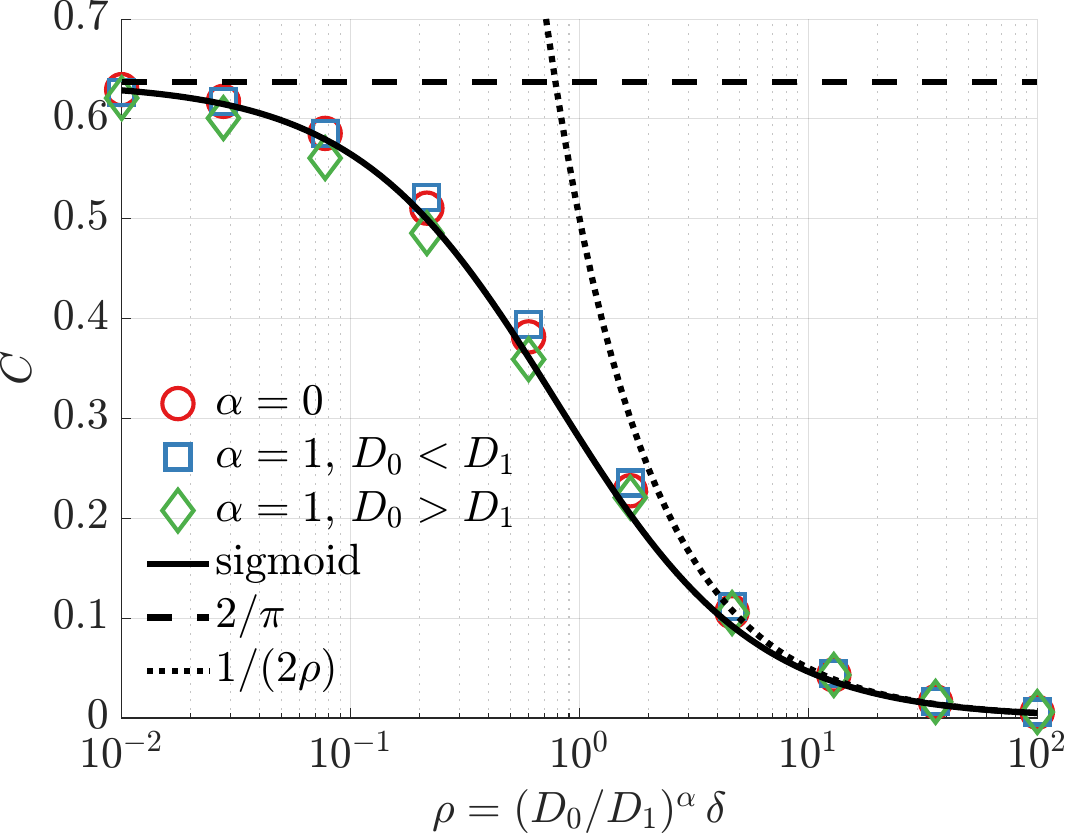}
    \caption{The markers show estimates of $C$ in \eqref{eq:corefar} computed from KMC simulations for the case that $\Gamma$ is the unit disk. The solid curve is the sigmoid approximation in \eqref{eq:sigmoiddisk}.}
    \label{fig:CK}
\end{figure}

Figure~\ref{fig:CK} plots estimates of the monopole $C$ in \eqref{eq:corefar} computed from KMC simulations for the case that $\Gamma$ is the unit disk. Each marker is the result of $10^5$ independent stochastic paths. The details of the stochastic simulation algorithm are in the Appendix. 

The markers labeled ``${\alpha}=0$'' in Figure~\ref{fig:CK} are for It\^{o} multiplicative noise, and the values of $D_0$ and $D_1$ are irrelevant. The markers labeled ``${\alpha}=1$'' in Figure~\ref{fig:CK} are for isothermal multiplicative noise, and for either $D_0/D_1=10$ or $D_0/D_1=1/10$. 

The solid curve is the sigmoid approximation in \eqref{eq:sigmoiddisk}, which agrees well with the simulations. For the case of a partially reactive trap with reactivity $\kappa$ (rather than a ``buried'' trap at depth $\delta$), the sigmoid formula in \eqref{eq:sigmoiddisk} with $(D_0/D_1)^{\alpha}\delta$ replaced by $1/\kappa$ was previously posited in \cite{berez04}, and a similar interpolation formula was suggested by Zwanzig and Szabo \cite{zwanzig1991time}. 

\section{Narrow escape through a tube}\label{sec:mfpt}

We now combine our analysis in section~\ref{sec:core} with matched asymptotics to determine the narrow escape time through a tube.

\subsection{Constructing the domain}
Let $\Omega_0'\subset\R^3$ be a bounded, three-dimensional domain with a smooth boundary $\partial\Omega_0'$. We now attach a ``tube'' $\Omega_1'$ to the ``bulk'' $\Omega_0'$. Let $\x_1\in\partial\Omega_0'$ and choose our Cartesian coordinate system $(x,y,z)\in\R^3$ so that the origin is $\x_1=(0,0,0)$, with $x$ along the normal vector from $\x_1$ (with $x<0$ pointing outside $\Omega_0'$) and $(y,z)\in\R^2$ specifying points in the tangent plane to $\partial\Omega_0'$ at $\x_1$. Let $\Gamma\in\R^2$ be a simply connected open set containing the origin with diameter $\textup{diam}(\Gamma)=2>0$. Let
\begin{align*}
    a\Gamma:=\{(y,z)\in\R^2:(y/a,z/a)\in\Gamma\}
\end{align*}
be a rescaling of $\Gamma$ so that $\text{diam}(a\Gamma)=2a$. Define
\begin{align*}
    \Omega_1'
    :=\{(x,y,z)\in\R^3:x\in(-L,\xi),(y,z)\in a\Gamma\},
\end{align*}
for some $L>0$, where $\xi\ge0$ is such that
\begin{align}\label{eq:xi}
    \{(x,y,z)\in\R^3:x=\xi,(y,z)\in a\Gamma\}\subset\Omega_0'.
\end{align}
In words, $\xi\ge0$ in \eqref{eq:xi} is large enough so that the tube $\Omega_1'$ sticks into $\Omega_0'$, but not so large that it pokes out of $\Omega_0'$. 
Note that we can always take $a>0$ small enough to ensure that \eqref{eq:xi} holds for some $\xi\ge0$. 
Assume that $\Omega_0'\cap\Omega_1'$ is simply connected for all $a>0$, which ensures that the tube $\Omega_1'$ does not intersect the domain $\Omega_0'$ in more than one place.

Finally, define a new ``tube'' $\Omega_1$ and a new ``bulk'' $\Omega_0$ so that their interface is flat and lies in the $x=0$ plane,
\begin{align}\label{eq:flat}
\begin{split}
    \Omega_1
    &:=\Omega_1'\backslash\{(x,y,z)\in\R^3:x\ge0\},\\
    \Omega_0
    &:=\{\Omega_0'\cup\Omega_1'\}\backslash\Omega_1.
\end{split}    
\end{align}

\subsection{Diffusion process}
Consider a path $\X=\{\X(t)\}_{t\ge0}$ of a particle that diffuses in the union of the bulk and the tube, $\Omega:=\Omega_0\cup\Omega_1$, according to the following stochastic differential equation written in terms of its infinitesimal increments
\begin{align}\label{eq:sdefull}
    \X(t+\dd t)
    = \X(t)+\sqrt{2D(\X_*)}\,\dd \mathbf{W},
\end{align}
where $D:\R^3\mapsto(0,\infty)$ is a given space-dependent diffusivity, $\mathbf{W}=\{\mathbf{W}(t)\}_{t\ge0}$ is a standard three-dimensional Brownian motion, and $\X_*$ interpolates between $\X(t)$ and $\X(t+\dd t)$,
\begin{align*}
    \X_*
    =(1-{\alpha})\X(t)+{\alpha} \X(t+\dd t),
\end{align*}
for a given multiplicative noise parameter ${\alpha}\in[0,1]$. Recall that ${\alpha}=0$, ${\alpha}=1/2$, and ${\alpha}=1$ correspond respectively to the It\^{o}, Stratonovich, and isothermal interpretations of the multiplicative noise in \eqref{eq:sdefull}. Suppose the particle has absorbing boundary conditions at the bottom of the tube,
\begin{align*}
    \partial\Omega_{a}
    :=\{(x,y,z)\in\R^3:x=-L,(y,z)\in a\Gamma\},
\end{align*}
and reflecting boundary conditions on the rest of the boundary, $\partial\Omega_r=\partial\Omega\backslash\partial\Omega_a$.

Let $\tau$ be the random absorption time,
\begin{align*}
    \tau
    :=\inf\{t\ge0:\X(t)\in\partial\Omega_a\}
    =\tau_0+\tau_1,
\end{align*}
which we have decomposed into the respective residence times in $\Omega_0$ and $\Omega_1$,
\begin{align*}
    \tau_0
    &:=\int_0^\tau 1_{\X(t)\in\Omega_0}\,\dd t,\quad 
    \tau_1
    :=\int_0^\tau 1_{\X(t)\in\Omega_1}\,\dd t,
\end{align*}
where $1_A$ denotes the indicator function on the event $A$ (i.e.\ $1_A=1$ if $A$ occurs and $1_A=0$ otherwise). 

Let $v(x)$ denote the mean residence time in $\Omega_0$ conditioned on the starting position $X(0)=\x$,
\begin{align*}
    v(\x)
    =\E[\tau_0\,|\,\X(0)=\x].
\end{align*} 
The mean residence time satisfies the following Poisson equation with mixed Dirichlet-Neumann boundary conditions \cite{vaccario2015, bressloff2017temporal},
\begin{align}
\begin{split}\label{eq:pdemfpt0}
D^{1-{\alpha}}\nabla\cdot[D^{\alpha}\nabla v]&=-1_{\x\in\Omega_0} , \quad \x \in \Omega,\\
\partial_n v & = 0,\quad \x\in\partial\Omega_r,\\
v & = 0,\quad \x\in\partial\Omega_a,
\end{split}
\end{align}
where $\partial_n$ denotes the normal derivative.

Suppose that the space-dependent diffusion coefficient $D(x)$ is $D_0$ in the bulk and $D_1$ in the tube,
\begin{align*}
    D(x)
    =\begin{cases}
        D_0 & \text{if }x\in\Omega_0,\\
        D_1 & \text{if }x\in\Omega_1.
    \end{cases}
\end{align*}
Hence, the boundary value problem \eqref{eq:pdemfpt0} becomes
\begin{align}
\begin{split}\label{eq:pdemfpt}
    D_0\Delta v
    & = -1_{\mathbf{x}\in\Omega_0},\quad \x\in\Omega,\\
    \partial_n v & = 0,\quad \x\in\partial\Omega_r,\\
v & = 0,\quad \x\in\partial\Omega_a,
\end{split}
\end{align}
with the condition that $v(\x)$ and $D(\x)^{\alpha}\nabla v(\x)$ are continuous at the interface of the bulk and the tube,
\begin{align}\label{eq:continuityv}
\begin{split}
    v(0^+,y,z)
    &=v(0^-,y,z),\quad (y,z)\in a\Gamma,\\
    D_0^{\alpha} \partial_x v(0^+,y,z)
    &=D_1^{\alpha} \partial_x v(0^-,y,z),\quad (y,z)\in a\Gamma.
    \end{split}
\end{align}

\subsection{Tube residence time}

Before studying the residence time $\tau_0$ in the bulk $\Omega_0$, we first consider the residence time $\tau_1$ in the tube $\Omega_1$. 
Since we defined $\Omega_0$ and $\Omega_1$ in \eqref{eq:flat} so that their interface is flat, it follows that the residence time $\tau_1$ in the tube $\Omega_1$ can be obtained by studying a mere one-dimensional diffusion process.

In particular, $\tau_1$ is equal in distribution to the absorption time of a one-dimensional diffusion process $\widetilde{X}$ in the interval $[-L,0]$ with diffusivity $D_1>0$, an absorbing boundary condition at $-L$, and a reflecting boundary condition at $0$. The initial position of $\widetilde{X}$ is either $0$ if $\X(0)\in\Omega_0$ or $x$ if $\X(0)=(x,y,z)\in\Omega_1$. Hence, it is straightforward to obtain the full probability distribution of $\tau_1$. However, for our purposes, we need only the mean residence time, which is
\begin{align}\label{eq:meantau1}
    \E[\tau_1\,|\,\X(0)=\x=(x,y,z)]
    =\begin{cases}
       L^2/(2D_1) & \text{if }\x\in\Omega_0,\\
       (L-|x|)(L+|x|)/(2D_1) & \text{if }\x\in\Omega_1.
    \end{cases}
\end{align}
We emphasize that \eqref{eq:meantau1} is exact.

\subsection{Matched asymptotic analysis}

We now use matched asymptotic analysis to study the solution to \eqref{eq:pdemfpt} as the tube radius vanishes. The analysis follows \cite{cheviakov2010asymptotic} and is related to the strong localized perturbation analysis pioneered in \cite{ward1993}. 

We first nondimensionalize the problem by rescaling space by the lengthscale of the bulk domain, $|\Omega_0|^{1/3}$, so that the tube now has cross-section diameter $\text{diam}(\eps\Gamma)=2\eps>0$ and length $L/|\Omega_0|^{1/3}=\eps L/a>0$, where
\begin{align*}
    \eps=a/|\Omega_0|^{1/3}\ll1.
\end{align*}
We further rescale time so that the particle has unit diffusivity in the bulk domain and diffusivity $D_1/D_0$ in the tube.

In the region in $\Omega_0$ away from an $\mathcal{O}(\eps)$ neighborhood of the tube, we expand the outer solution as
\begin{align}
v\sim \eps^{-1}v_0+v_1 + \cdots,\label{eq:out_sol1}
\end{align}
where $v_0$ is an unknown constant and $v_1$ is a function of $\x$. Substituting \eqref{eq:out_sol1} into \eqref{eq:pdemfpt} and matching the $\mathcal{O}(1)$ terms yields
\begin{alignat}{2}
\Delta v_1 &= -1, \quad &&\x \in \Omega_0, \label{eq:v1a}\\
\partial_n v_1&=0, \quad &&\x \in \partial\Omega_0\backslash\x_1.\label{eq:v1b}
\end{alignat}
Notice that from the perspective of the outer solution, the tube has shrunk to a point.

In the region near the tube, we introduce the inner variables around $\x_1=(0,0,0)$,
\begin{align*}
\x'
&:=\eps^{-1}(\x-\x_1)
=\eps^{-1}\x,\\
w(\x')
&:=v(\x_1+\eps\x')
=v(\eps\x').
\end{align*}
Applying the Laplacian in $\x'$ to the inner solution expansion $w\sim \eps^{-1}w_0$ yields
 \begin{align*}
    \Delta w_0
    &=0, \quad \{(x',y',z')\in\R^3:x'>0\}\cup\{(x',y',z')\in[0,-\delta)\times\Gamma\},\\
    w_0
    &=0, \quad x'=-L/a,\,(y',z')\in\Gamma,\\
    \partial_{x'} w_0
    &=0, \quad x'=0,\,(y',z')\notin\overline{\Gamma},\\
    \partial_{n}w_0
    &=0,\quad x'\in(-L/a,0),\,(y',z')\in\partial\Gamma,
\end{align*}
where $\Gamma$ has diameter $\text{diam}(\Gamma)=2$. Furthermore, the continuity conditions on $v$ in \eqref{eq:continuityv} imply that $w_0$ satisfies the continuity conditions in \eqref{eq:cont2}. 
The matching condition ensures that the near-field behavior of the outer solution as $\x \to \x_1$ agrees with the far-field behavior of the inner solution,
\begin{align}
\eps^{-1}v_0+v_1 + \cdots
\sim \eps^{-1}w_0+\cdots.
\label{eq:matching}
\end{align}
Therefore, if $u$ solves \eqref{eq:core}-\eqref{eq:cont2} with depth $\delta=L/a$, then
\begin{align}\label{eq:wcbc}
    w_0
    =v_0(1-u).
\end{align}

The far-field behavior in \eqref{eq:corefar}, the relation \eqref{eq:wcbc}, and the matching condition \eqref{eq:matching} yields the following singularity condition on $v_1$, 
\begin{align}\label{eq:sing}
v_1
\sim -\frac{v_0C(L/a,\Gamma,(D_0/D_1)^{\alpha})}{\|\x-\x_n\|}\quad \text{ as }\x \to \x_1.
\end{align}
Using the divergence theorem, the solvability condition for \eqref{eq:v1a}-\eqref{eq:v1b} with the singularity in \eqref{eq:sing} is $-1=-2\pi v_0 C$, and therefore 
\begin{align}\label{eq:v0}
v_0
=\frac{1}{2\pi  C(L/a,\Gamma,(D_0/D_1)^{\alpha})}.
\end{align}
Recalling that $v\sim\eps^{-1}v_0$ from \eqref{eq:out_sol1}, the leading order behavior of the (dimensionless) expected residence time $\tau_0$ in $\Omega_0$ is
\begin{align}\label{eq:mfptdimless}
v(\x)
=\E[\tau_0\,|\,\X(0)=\x]
\sim\frac{1}{2\pi\eps C(L/a,\Gamma,(D_0/D_1)^{\alpha})}\quad\text{as }\eps\to0,
\end{align}
which is valid for initial positions $\x\in\Omega_0$ which are outside of an $\mathcal{O}(\eps)$ neighborhood of the entrance to the tube. 

\subsection{Escape time}

We now use the analysis above to investigate the mean escape time. Putting the bulk residence time \eqref{eq:mfptdimless} in dimensional form (i.e.\ dividing by $D_0/|\Omega_0|^{2/3}$ and recalling that $\eps=a/|\Omega_0|^{1/3}$) and combining with the tube residence time in \eqref{eq:meantau1} yields the following mean escape time asymptotic,
\begin{align}\label{eq:mfpt}
    \E[\tau]
    \sim\frac{|\Omega_0|}{2\pi a D_0}\frac{1}{C(L/a,\Gamma,(D_0/D_1)^{\alpha})}
    +\frac{L^2}{2D_1}\quad\text{as }a/|\Omega_0|^{1/3}\to0,
\end{align}
which is valid for any starting location $\x\in\Omega_0$ whose distance from the tube is much greater than the tube cross-section diameter (i.e.\ $\|\x\|\gg\text{diam}(a\Gamma)=2a$).

We now investigate \eqref{eq:mfpt} using the analysis of $C=C(L/a,\Gamma,(D_0/D_1)^{\alpha})$ from section~\ref{sec:core}. By \eqref{eq:to0},
\begin{align}\label{eq:mfpt1}
    \E[\tau]
    \approx\frac{|\Omega_0|}{2\pi a D_0}\frac{1}{C_0(\Gamma)}
    +\frac{L^2}{2D_1}\quad\text{if $a/|\Omega_0|^{1/3}\ll1$ and $(L/a)(D_0/D_1)^{\alpha}\ll1$},
\end{align}
where $C_0(\Gamma)$ is the electrostatic capacitance of $\Gamma$. By \eqref{eq:to1},
\begin{align}\label{eq:mfpt2}
    \E[\tau]
    &\approx\frac{|\Omega_0|L}{|\Gamma|a^2 D_0^{1-\alpha}D_1^\alpha}
    +\frac{L^2}{2D_1}\quad\text{if $a/|\Omega_0|^{1/3}\ll1$ and $(L/a)(D_0/D_1)^{\alpha}\gg1$}.
\end{align}
Using the sigmoid approximation to $C$ in \eqref{eq:sigmoid}, 
\begin{align}\label{eq:mfpt3}
        \E[\tau]
    &\approx\frac{|\Omega_0|}{2\pi a D_0}\frac{1}{C_0(\Gamma)}
    +\frac{|\Omega_0|L}{|\Gamma|a^2 D_0^{1-\alpha}D_1^\alpha}
    +\frac{L^2}{2D_1}\quad\text{if $a/|\Omega_0|^{1/3}\ll1$}.
\end{align}
Note that if $\Gamma$ is the unit disk (i.e.\ the tube is a cylinder with radius $a>0)$, then $C_0(\Gamma)=2/\pi$ and $|\Gamma|=\pi$.

The results in \eqref{eq:mfpt1}-\eqref{eq:mfpt3} show the importance of the parameter $\alpha\in[0,1]$ describing the multiplicative noise inherent to a space-dependent diffusivity. For instance, the residence time in the bulk is independent of the tube diffusivity $D_1$ if and only if $\alpha=0$ (i.e. if any only if the noise is It\^{o}). Furthermore, if $\alpha=1$ (sometimes called isothermal noise), then the escape time is independent of the bulk diffusivity $D_0$ in the regime in \eqref{eq:mfpt2}. Furthermore, if $\alpha\neq0$, then decreasing the tube diffusivity $D_1$ can shift the escape time from \eqref{eq:mfpt1} to \eqref{eq:mfpt2}, whereas if $\alpha=0$, then only the tube aspect ratio $L/a$ determines the validity of \eqref{eq:mfpt1} versus \eqref{eq:mfpt2}.


\section{Discussion}\label{sec:discussion}

In this paper, we determined the exact asymptotics of the diffusive escape time through a tube. For a cylindrical tube, our results can be most succinctly stated with the single formula,
\begin{align}\label{eq:ours}
    \E[\tau]
    &\approx\frac{|\Omega_0|L}{\pi a^2 D_0^{1-\alpha}D_1^\alpha}
    +\frac{|\Omega_0|}{4 D_0a}
    +\frac{L^2}{2D_1}\quad\text{if }a/|\Omega_0|^{1/3}\ll1,
\end{align}
which gives the exact asymptotic escape time if $\rho\ll1$ or $\rho\gg1$, where $\rho:=(L/a)(D_0/D_1)^\alpha$. If $D_0=D_1$ and $\rho\gg1$, then the first term in \eqref{eq:ours} reduces to the time in \eqref{eq:T1} suggested in \cite{svoboda1996direct} from electrical analogies and used in \cite{svoboda1996direct, majewska2000regulation, bloodgood2005neuronal} to interpret neuron activity. In this regime of $D_0=D_1$ and $\rho\gg1$, our estimate \eqref{eq:ours} is identical to the estimate \eqref{eq:T3}, except for a numerical prefactor in the first term that differs by around 7\% (\eqref{eq:T3} was obtained by fitting a functional form to numerical simulations). If $D_0=D_1$ and $\rho\ll1$, then \eqref{eq:ours} reduces to \eqref{eq:T2}, which was derived by assuming that the particle cannot return to the bulk after entering the tube. If $D_0\neq D_1$, then our estimate \eqref{eq:ours} differs from the estimate \eqref{eq:T4} derived in \cite{berezhkovskii2009escape} unless $\alpha=1$. Taking $\alpha=1$ is termed the isothermal (or kinetic or Hanggi-Klimontovich) interpretation of multiplicative noise \cite{hanggi1982stochastic, klimontovich1990ito, ao2007existence} and is known to produce counterintuitive first passage time predictions \cite{vaccario2015, tung2025escape, tung2026stochastic}.

Our results show that escape through a tube is similar to absorption at a partially absorbing (or partially reactive) trap \cite{chaigneau2022first, grebenkov2017effects, grebenkov2017diffusive, grebenkov2019full, guerin2023imperfect}. To explain, if the tube is replaced by a partially absorbing disk with radius $a>0$ and reactivity $\kappa>0$, then the mean absorption time can be approximated by \cite{chaigneau2022first, grebenkov2025asymptotic}
\begin{align}\label{eq:kappa}
    \E[\tau_{\kappa}]
    \approx\frac{|\Omega_0|}{\pi a^2\kappa}
    +\frac{|\Omega_0|}{4D_0a}.
\end{align}
The estimate \eqref{eq:kappa} is identical to the first two terms in \eqref{eq:ours} (which give the mean residence time in the bulk $\Omega_0$) if we take the reactivity parameter to be
\begin{align*}
    \kappa
    =\frac{D_0^{1-\alpha}D_1^\alpha}{L}>0.
\end{align*}
Naturally, a long tube (large $L$) is akin to low reactivity (small $\kappa$), and vice versa.

Finally, as described in the Introduction section, the escape time through a tube is relevant to understanding the mechanism(s) underlying the compartmentalization observed in budding yeast asymmetric protein segregation. The dumbbell shape of the anaphase nucleus of budding yeast can be approximated by two spheres connected by a cylindrical tube, where the volume of each sphere and the length and radius of the tube are respectively \cite{boettcher2012nuclear}
\begin{align}\label{eq:yeast}
    |\Omega_0|
    =\tfrac{4}{3}\pi(0.8)^3\,\mu\mathrm{m}^3
    \approx2.1\,\mu\mathrm{m}^3,\quad
    L
    =2.5\,\mu\mathrm{m},\quad
    a
    =0.15\,\mu\mathrm{m}.
\end{align}
Given estimates of the diffusivity of a protein (i.e.\ $D_0$, $D_1$, and $\alpha$), the formula \eqref{eq:ours} predicts the time it takes that protein to diffuse from the mother nuclear lobe to the daughter nuclear lobe (and vice versa). It is difficult to directly compare such predicted times to experimental data, since the relevant experimental data is typically described in terms of fluorescence decay rates from fluorescence loss in photobleaching (FLIP) experiments. Nevertheless, \eqref{eq:ours} predicts how this time depends on nuclear envelope geometry, and these predictions agree with some prior estimates. Specifically, using the parameters in \eqref{eq:yeast}, simulated FLIP experiments found that increasing the tube radius $a$ by a factor of 3.4 ``decreased the compartmentalization of TetR-GFP approximately eightfold'' \cite{boettcher2012nuclear}, where the degree of compartmentalization was estimated by comparing the fluorescence decay rates in the mother and daughter nuclear lobes, which is ``inversely proportional to the exchange rate between the two compartments'' \cite{boettcher2012nuclear} (and is thus proportional to the escape time $\E[\tau]$). 
If we plug these parameters into \eqref{eq:ours} and increase $a$ by a factor of 3.4, then $\E[\tau]$ decreases by a factor of 7.8, which agrees closely with the aforementioned eightfold decrease found in simulations \cite{boettcher2012nuclear}. Similarly, dividing $L$ by 3 ``reduced compartmentalization $\sim2.5$-fold'' in simulations \cite{boettcher2012nuclear}, and the formula \eqref{eq:ours} predicts that this decrease in $L$ decreases $\E[\tau]$ by a factor of 2.8. Finally, replacing $a$ by $3.4a$ and $L$ by $L/3$ ``abolished compartmentalization'' in simulations \cite{boettcher2012nuclear}, and these geometric changes decrease \eqref{eq:ours} by a factor of 23.

\appendix
\section{KMC simulation algorithms}

We now describe the KMC simulation algorithms we use to numerically approximate the monopole $C$ in \eqref{eq:corefar}. Our simulation algorithm supposes that $\Gamma$ is the unit disk,
\begin{align*}
    \Gamma
    =\{(y,z)\in\R^2:y^2+z^2<1\}.
\end{align*}
In brief, we simulate paths $(X,Y,Z)$ of many diffusing particles and count the fraction that get absorbed before reaching a large distance from the absorbing trap. This fraction approximates $u$ in section~\ref{sec:probinterp}, from which we approximate $C$ via \eqref{eq:rela}. Each marker is for $10^5$ independent trials. More precisely, the algorithm takes the following steps.

\textbf{Step 1:} Start a particle uniformly distributed on the surface of the unit hemisphere centered at the origin (i.e.\ on the surface of $H_R$ in \eqref{eq:hemi} with $R=1$).

\textbf{Step 2:} Following the algorithm devised by Bernoff, Lindsay, and Schmidt \cite{bernoff_boundary_2018}, simulate the path of the particle until it either (i) hits the intersection of $\Gamma$ with the $x=0$ plane or (ii) reaches a distance $R_\infty\gg1$ from the origin (we take $R_\infty=10^{10}$). If (ii) occurs, then we assume that the particle will never get absorbed (its probability of getting absorbed is on the order of $1/R_\infty$ if it reaches distance $R_\infty\gg1$) and end the simulation. If (i) occurs, then we proceed to Step 3. The specifics of Step 3 depend on the values of $D_0$, $D_1$, and/or ${\alpha}\in[0,1]$.

\textbf{Step 3 for approximating $C$ when $D_0=D_1$:}
To simplify the exposition, set $t=0$. Sample the time $\tau_{\text{disk}}$, which is the first time that $(Y,Z)$ hits the boundary of the unit disk $\Gamma$. Letting $r_0=\sqrt{Y^2(0)+Z^2(0)}<1$, a straightforward separation of variables calculation yields that the cumulative distribution function of $\tau_{\text{disk}}$ is
\begin{align}\label{eq:taudisk}
    F(t;r_0)
    =\P(\tau_{\text{disk}}\le t\,|\,Y^2(0)+Z^2(0)=r_0^2)
    =1-\sum_{n=1}^\infty \frac{2J_0(a_n r_0)}{a_n J_1(a_n)}e^{-Da_n^2t},
\end{align}
where $D=D_0=D_1$, $J_0$ and $J_1$ are the Bessel functions of the first kind of respective order 0 and 1, and $a_n$ is the $n$th positive zero of $J_0$ (i.e.\ $J_0(a_n)=0$). The realization $\tau_{\text{disk}}$ is then sampled by numerically solving $F(\tau_{\text{disk}};r_0)=U$, where $U\in(0,1)$ is uniformly distributed on $(0,1)$.

On the time interval $(0,\tau_{\text{disk}})$, we are assured that $(Y,Z)\in\Gamma$ and $X$ is a one-dimensional diffusion process on the semi-infinite interval $(-\delta,\infty)$ with an absorbing boundary condition at $x=-\delta<0$ and initial position $X(0)=0$. Define the shifted process $\widehat{X}=X+\delta$ (which thus begins at $\widehat{X}(0)=\delta>0$ and gets absorbed if $\widehat{X}=0$), and consider the following partial cumulative distribution function of $\widehat{X}(t)$,
\begin{align}\label{eq:Fcdf}
\begin{split}
    F(x;x_0,t)
    &:=\P(\widehat{X}(t)\le x\,|\,\widehat{X}(0)=x_0)\\
    &\;=\frac{1}{2}\bigg[\textup{erf}\Big(\frac{x-x_0}{\sqrt{4Dt}}\Big)-\textup{erf}\Big(\frac{x+x_0}{\sqrt{4Dt}}\Big)+2\textup{erf}\Big(\frac{x_0}{\sqrt{4Dt}}\Big)\bigg].
\end{split}        
\end{align}
The formula~\ref{eq:Fcdf} can be obtained via the method of images \cite{carslaw1959}. 
The probability that $X$ does not hit get absorbed before time $\tau_{\text{disk}}$ is
\begin{align*}
    P_{\text{survive}}
    :=\lim_{x\to\infty}F(x;\delta,\tau_{\text{disk}})
    =\text{erf}(\delta/\sqrt{4D\tau_{\text{disk}}}).
\end{align*}
Let $U\in(0,1)$ be another independent uniform random variable. If $U>P_{\text{survive}}$, then the particle got absorbed and the simulation ends. Otherwise, we numerically solve $F(\widehat{X}(\tau_{\text{disk}});\delta,\tau_{\text{disk}})=U$ for $\widehat{X}(\tau_{\text{disk}})$ and set $X(\tau_{\text{disk}})=\widehat{X}(\tau_{\text{disk}})-\delta$.

If $X(\tau_{\text{disk}})>0$ (i.e.\ the particle is above the pit), then we set $(Y(\tau_{\text{disk}}),Z(\tau_{\text{disk}}))=(1,0)$ (note that we assured that $Y^2(\tau_{\text{disk}})+Z^2(\tau_{\text{disk}})=1$, and rotating about the $y=z=0$ line does not affect the fate of the particle), and we return to Step 2 above.

\textbf{Step 3b:} If $X(\tau_{\text{disk}})<0$ (i.e.\ the particle is in the pit), then we set $t=0$ and $x_0=X(\tau_{\text{disk}})<0$. We need to determine if the particle gets absorbed at $x=-\delta$ before reaching $x=0$. Consider the partial cumulative distribution function \cite{linn_extreme_2022}
\begin{align*}
    \Phi(s,w)
    =\begin{cases}\sum_{k=1}^{\infty}(1-e^{-k^{2}\pi^{2}s})\frac{2}{k\pi}\sin(k\pi w),\\
\sum_{k=-\infty}^{\infty}\textup{sgn}(2k+w)\textup{erfc}\Big(\frac{|2k+w|}{\sqrt{4s}}\Big).
\end{cases}
\end{align*}
To explain $\Phi$, let $\overline{X}$ be a one-dimensional diffusion process in the interval $(0,1)$ with unit diffusivity, initial position $\overline{X}(0)=w\in(0,1)$, and absorbing boundary conditions at both $x=0$ and $x=1$. If $\tau_{\text{right}}=\inf\{t\ge0:\overline{X}(t)=1\}$ is the first time that $\overline{X}$ hits $x=1$ (with $\tau_{\text{right}}=+\infty$ if $\overline{X}$ hits $x=0$), then $\Phi$ is the partial cumulative distribution function of $\tau_{\text{right}}$, where the two expressions for $\Phi$ are its large-time and small-time expansions (i.e.\ the first expression converges rapidly for large time $s$ and the second expression converges rapidly for small time $s$). The probability that $\tau_{\text{right}}=+\infty$ is $\lim_{s\to\infty}\Phi(s,w)=1-w$. Let $U\in(0,1)$ be another independent uniform random variable and set $w=x_0/\delta$. If $U>1-w$, then the particle got absorbed and the simulation ends. Otherwise, we sample the time $\tau_{\text{top}}$ by numerically solving $\Phi(D\tau_{\text{top}}/\delta^2,w)=U$ and we set $X(\tau_{\text{top}})=0$. To determine $(Y(\tau_{\text{top}}),Z(\tau_{\text{top}}))$, we note that on the time interval $(0,\tau_{\text{top}})$, the coordinates $(Y,Z)$ diffuse in a unit disk with reflecting boundary conditions. Letting $r_0=\sqrt{Y^2(0)+Z^2(0)}<1$, a straightforward separation of variables calculation shows that the probability density function for the radial position of such a process at time $t\in(0,\tau_{\text{top}})$ is
\begin{align*}
    p(r,t;r_0)
    =2+\sum_{n=1}^\infty \frac{J_0(\beta_n r_0)}{\int_0^1 J_0^2(\beta_n s) s\,\dd s}J_0(\beta_n r) e^{-D\beta_n^2t},
\end{align*}
where $\beta_n$ is the $n$th positive zero of $J_1$, i.e.\ $J_1(\beta_n)=0$. Since the corresponding cumulative distribution function is
\begin{align*}
    G(r,t;r_0)
    =\P(Y^2(t)+Z^2(t)\le r^2\,|\,Y^2(0)+Z^2(0)=r_0^2)
    =\int_0^r p(r',t;r_0)\,r'\,\dd r',
\end{align*}
we sample $Y(\tau_{\text{top}})$ by numerically solving $G(Y(\tau_{\text{top}}),\tau_{\text{top}};r_0)=U$ where $U\in(0,1)$ is another independent uniform random variable, and set $Z(\tau_{\text{top}})=0$ (again, recall that rotating along the $y=z=0$ line does not affect the fate of the particle). We then proceed to Step 2 above.

\textbf{Step 3 for approximating $C$ when $D_0\neq D_1$ and ${\alpha}=0$:} 
If ${\alpha}=0$, then the particle simply moves faster or slower depending on if $X>0$ or $X<0$ and the values of $D_0$ and $D_1$. However, the ultimate fate of the particle (i.e.\ getting absorbed or not) is not affected by such a time change. Hence, Step 3 for the case ${\alpha}=0$ is the same as Step 3 for the case $D_0=D_1$ described above.

\textbf{Step 3 for approximating $C$ when $D_0\neq D_1$ and ${\alpha}\neq0$:} 
If $D_0\neq D_1$ and ${\alpha}\neq0$, then there is an effective ``force'' at $x=0$ that either pushes the particle up if $D_0>D_1$ (i.e.\ larger diffusivity above the pit) or down if $D_0<D_1$ (i.e.\ smaller diffusivity above the pit). Introduce a small simulation parameter $\eta>0$ and set $t=0$ so that $X(0)=0$ and $Y^2(0)+Z^2(0)<1$ at the start of this step. In our simulations, we take $\eta=\min\{0.1,\delta/2\}$. As an approximation, we assume that $Y(t)=Y(0)$ and $Z(t)=Z(0)$ for all $t\in[0,\tau_{\text{esc}}]$, where $\tau_{\text{esc}}=\inf\{t\ge0:X(t)\notin(-\eta,\eta)\}$ is the first time that $X$ reaches $\pm\eta$. Let $P_{\text{up}}\in(0,1)$ be the probability that $X(\tau_{\text{esc}})=+\eta$, which we show below is 
\begin{align}\label{eq:musplit}
    P_{\text{up}}
    =\frac{D_0^{\alpha}}{D_0^{\alpha}+D_1^{\alpha}}.
\end{align}
With probability $P_{\text{up}}$, we set $(X(\tau_{\text{esc}}),Y(\tau_{\text{esc}}),Z(\tau_{\text{esc}}))=(+\eta,Y(0),Z(0))$ and go to Step 2 above. Otherwise, we proceed as in Step 3b above, which takes the particle either to absorption at $x=-\delta$ or to $x=0$. If the particle reaches $x=0$, then we return to the start of this Step 3.

To obtain \eqref{eq:musplit}, consider a one-dimensional diffusion process. Suppose that the particle has diffusivity $D_0$ if $x<L/2$ and diffusivity $D_1$ if $x>L/2$, where the change in diffusivity at $x=L/2$ is interpreted with multiplicative noise parameter ${\alpha}$. If $P(x_0)$ is the probability that the particle reaches $x=L$ before $x=0$ given that it starts at $x_0\in[0,L]$, then $P(x_0)$ satisfies
\begin{align*}
    P''=0,\quad\text{if }x\in(0,L/2)\cup(L/2,L),
\end{align*}
with continuity conditions at $L/2$,
\begin{align*}
    P(L/2-)
    &=P(L/2+),\\
    D_0^{\alpha} P'(L/2-)
    &=D_1^{\alpha} P'(L/2+),
\end{align*}
with boundary conditions $P(0)=0=1-P(L)$. Hence, 
\begin{align*}
    P(x_0)
    =\begin{cases}
        \frac{2D_1^{\alpha}}{D_0^{\alpha}+D_1^{\alpha}}\frac{x}{L} & \text{if }x\in[0,L/2],\\
        1-\frac{2D_0^{\alpha}}{D_0^{\alpha}+D_1^{\alpha}}+\frac{2D_0^{\alpha}}{D_0^{\alpha}+D_1^{\alpha}}\frac{x}{L} & \text{if }x\in[L/2,L].
    \end{cases}
\end{align*}
Hence, $P_{\text{up}}=1-P(L/2)$, which yields \eqref{eq:musplit}.

\bibliography{library.bib}

@article{tung2026stochastic,
  title={Stochastic search with space-dependent diffusivity},
  author={Tung, Hwai-Ray and Lawley, Sean D},
  journal={arXiv preprint arXiv:2601.08740},
  year={2026}
}

@article{tung2025escape,
  title={Escape from heterogeneous diffusion},
  author={Tung, Hwai-Ray and Lawley, Sean D},
  journal={arXiv preprint arXiv:2512.19646},
  year={2025}
}

@article{zavala2014long,
  title={The long and viscous road: uncovering nuclear diffusion barriers in closed mitosis},
  author={Zavala, Eder and Marquez-Lago, Tatiana T},
  journal={PLoS computational biology},
  volume={10},
  number={7},
  pages={e1003725},
  year={2014},
  publisher={Public Library of Science San Francisco, USA}
}

@article{shcheprova2008mechanism,
  title={A mechanism for asymmetric segregation of age during yeast budding},
  author={Shcheprova, Zhanna and Baldi, Sandro and Frei, Stephanie Buvelot and Gonnet, Gaston and Barral, Yves},
  journal={Nature},
  volume={454},
  number={7205},
  pages={728--734},
  year={2008},
  publisher={Nature Publishing Group UK London}
}

@article{boettcher2012nuclear,
  title={Nuclear envelope morphology constrains diffusion and promotes asymmetric protein segregation in closed mitosis},
  author={Boettcher, Barbara and Marquez-Lago, Tatiana T and Bayer, Mathias and Weiss, Eric L and Barral, Yves},
  journal={Journal of Cell Biology},
  volume={197},
  number={7},
  pages={921--937},
  year={2012},
  publisher={The Rockefeller University Press}
}

@article{grebenkov2025asymptotic,
  title={The Asymptotic Analysis of Some PDE and Steklov Eigenvalue Problems with Partially Reactive Patches in 3-D},
  author={Grebenkov, Denis S and Ward, Michael J},
  journal={arXiv preprint arXiv:2509.17394},
  year={2025}
}

@article{chaigneau2022first,
  title={First-passage times to anisotropic partially reactive targets},
  author={Chaigneau, Adrien and Grebenkov, Denis S},
  journal={Physical Review E},
  volume={105},
  number={5},
  pages={054146},
  year={2022},
  publisher={APS}
}

@article{ao2007existence,
  title={On the existence of potential landscape in the evolution of complex systems},
  author={Ao, Ping and Kwon, Chulan and Qian, Hong},
  journal={Complexity},
  volume={12},
  number={4},
  pages={19--27},
  year={2007},
  publisher={Wiley Online Library}
}

@article{klimontovich1990ito,
  title={Ito, Stratonovich and kinetic forms of stochastic equations},
  author={Klimontovich, Yu L},
  journal={Physica A: Statistical Mechanics and its Applications},
  volume={163},
  number={2},
  pages={515--532},
  year={1990},
  publisher={Elsevier}
}

@article{hanggi1982stochastic,
  title={Stochastic processes: Time evolution, symmetries and linear response},
  author={H{\"a}nggi, Peter and Thomas, Harry},
  journal={Physics Reports},
  volume={88},
  number={4},
  pages={207--319},
  year={1982},
  publisher={Elsevier}
}

@article{isaacson2016uniform,
  title={Uniform asymptotic approximation of diffusion to a small target: Generalized reaction models},
  author={Isaacson, Samuel A and Mauro, Ava J and Newby, Jay},
  journal={Physical Review E},
  volume={94},
  number={4},
  pages={042414},
  year={2016},
  publisher={APS}
}

@article{benichou2008narrow,
  title={Narrow-escape time problem: Time needed for a particle to exit a confining domain through a small window},
  author={B{\'e}nichou, O and Voituriez, R},
  journal={Physical review letters},
  volume={100},
  number={16},
  pages={168105},
  year={2008},
  publisher={APS}
}

@article{zhou1998theory,
  title={Theory of the diffusion-influenced substrate binding rate to a buried and gated active site},
  author={Zhou, Huan-Xiang},
  journal={The Journal of chemical physics},
  volume={108},
  number={19},
  pages={8146--8154},
  year={1998},
  publisher={American Institute of Physics}
}

@article{zhou2010diffusion,
  title={Diffusion-influenced transport of ions across a transmembrane channel with an internal binding site},
  author={Zhou, Huan-Xiang},
  journal={The journal of physical chemistry letters},
  volume={1},
  number={13},
  pages={1973--1976},
  year={2010},
  publisher={ACS Publications}
}

@article{samson1978diffusion,
  title={Diffusion-controlled reaction rate to a buried active site},
  author={Samson, Rene and Deutch, JM},
  journal={The Journal of Chemical Physics},
  volume={68},
  number={1},
  pages={285--290},
  year={1978},
  publisher={American Institute of Physics}
}

@article{berezhkovskii2011diffusion,
  title={Diffusion-influenced ligand binding to buried sites in macromolecules and transmembrane channels},
  author={Berezhkovskii, Alexander M and Szabo, Attila and Zhou, Huan-Xiang},
  journal={The Journal of chemical physics},
  volume={135},
  number={7},
  year={2011},
  publisher={AIP Publishing}
}

@article{grebenkov2019full,
  title={Full distribution of first exit times in the narrow escape problem},
  author={Grebenkov, Denis S and Metzler, Ralf and Oshanin, Gleb},
  journal={New Journal of Physics},
  volume={21},
  number={12},
  pages={122001},
  year={2019},
  publisher={IOP Publishing}
}

@article{grebenkov2017diffusive,
  title={Diffusive escape through a narrow opening: new insights into a classic problem},
  author={Grebenkov, Denis S and Oshanin, Gleb},
  journal={Physical Chemistry Chemical Physics},
  volume={19},
  number={4},
  pages={2723--2739},
  year={2017},
  publisher={Royal Society of Chemistry}
}

@article{grebenkov2017effects,
  title={Effects of the target aspect ratio and intrinsic reactivity onto diffusive search in bounded domains},
  author={Grebenkov, Denis S and Metzler, Ralf and Oshanin, Gleb},
  journal={New Journal of Physics},
  volume={19},
  number={10},
  pages={103025},
  year={2017},
  publisher={IOP Publishing}
}

@article{guerin2023imperfect,
  title={Imperfect narrow escape problem},
  author={Gu{\'e}rin, Thomas and Dolgushev, M and B{\'e}nichou, O and Voituriez, R},
  journal={Physical Review E},
  volume={107},
  number={3},
  pages={034134},
  year={2023},
  publisher={APS}
}

@article{stratonovich1966new,
  title={A new representation for stochastic integrals and equations},
  author={Stratonovich, RL},
  journal={SIAM Journal on Control},
  volume={4},
  number={2},
  pages={362--371},
  year={1966},
  publisher={SIAM}
}

@article{ito1944stochastic,
  title={Stochastic integral},
  author={It{\^o}, Kiyosi},
  journal={Proceedings of the Imperial Academy},
  volume={20},
  number={8},
  pages={519--524},
  year={1944},
  publisher={The Japan Academy}
}

@article{mannella2012ito,
  title={It{\^o} versus Stratonovich: 30 years later},
  author={Mannella, Riccardo and McClintock, Peter VE},
  journal={Fluctuation and Noise Letters},
  volume={11},
  number={01},
  pages={1240010},
  year={2012},
  publisher={World Scientific}
}

@article{van1981ito,
  title={It{\^o} versus stratonovich},
  author={Van Kampen, Nicolaas G},
  journal={Journal of Statistical Physics},
  volume={24},
  number={1},
  pages={175--187},
  year={1981},
  publisher={Springer}
}

@article{berezhkovskii2009escape,
  title={Escape from cavity through narrow tunnel},
  author={Berezhkovskii, Alexander M and Barzykin, Alexander V and Zitserman, Vladimir Yu},
  journal={The Journal of chemical physics},
  volume={130},
  number={24},
  year={2009},
  publisher={AIP Publishing}
}

@article{zwanzig1991time,
  title={Time dependent rate of diffusion-influenced ligand binding to receptors on cell surfaces},
  author={Zwanzig, Robert and Szabo, Attila},
  journal={Biophysical Journal},
  volume={60},
  number={3},
  pages={671--678},
  year={1991},
  publisher={Elsevier}
}

@article{cheviakov2010asymptotic,
  title={{An asymptotic analysis of the mean first passage time for narrow escape problems: Part II: The sphere}},
  author={Cheviakov, Alexei F and Ward, Michael J and Straube, Ronny},
  journal={Multiscale Modeling \& Simulation},
  volume={8},
  number={3},
  pages={836--870},
  year={2010},
  publisher={SIAM}
}

@article{singer2006narrow2,
  title={Narrow escape, Part II: The circular disk},
  author={Singer, Amit and Schuss, Zeev and Holcman, David},
  journal={Journal of statistical physics},
  volume={122},
  number={3},
  pages={465--489},
  year={2006},
  publisher={Springer}
}

@article{singer2006narrow,
  title={Narrow escape, part I},
  author={Singer, Amit and Schuss, Zeev and Holcman, David and Eisenberg, Robert S},
  journal={Journal of Statistical Physics},
  volume={122},
  number={3},
  pages={437--463},
  year={2006},
  publisher={Springer}
}

@article{holcman2004escape,
  title={Escape through a small opening: receptor trafficking in a synaptic membrane},
  author={Holcman, David and Schuss, Z},
  journal={Journal of Statistical Physics},
  volume={117},
  number={5},
  pages={975--1014},
  year={2004},
  publisher={Springer}
}

@article{biess2007diffusion,
  title={Diffusion in a dendritic spine: the role of geometry},
  author={Biess, A and Korkotian, Eduard and Holcman, David},
  journal={Physical Review E—Statistical, Nonlinear, and Soft Matter Physics},
  volume={76},
  number={2},
  pages={021922},
  year={2007},
  publisher={APS}
}

@article{bloodgood2005neuronal,
  title={Neuronal activity regulates diffusion across the neck of dendritic spines},
  author={Bloodgood, Brenda L and Sabatini, Bernardo L},
  journal={Science},
  volume={310},
  number={5749},
  pages={866--869},
  year={2005},
  publisher={American Association for the Advancement of Science}
}

@article{majewska2000regulation,
  title={Regulation of spine calcium dynamics by rapid spine motility},
  author={Majewska, Ania and Tashiro, Ayumu and Yuste, Rafael},
  journal={Journal of Neuroscience},
  volume={20},
  number={22},
  pages={8262--8268},
  year={2000},
  publisher={Society for Neuroscience}
}

@article{svoboda1996direct,
  title={Direct measurement of coupling between dendritic spines and shafts},
  author={Svoboda, Karel and Tank, David W and Denk, Winfried},
  journal={Science},
  volume={272},
  number={5262},
  pages={716--719},
  year={1996},
  publisher={American Association for the Advancement of Science}
}

@article{richardson2025quantifying,
  title={{Quantifying the transition from single file to Fickian diffusion}},
  author={Richardson, Victorya and Lawley, Sean D},
  journal={Journal of Chemical Physics},
volume={163},
number={19},
  year={2025}
}

@article{kaye2020,
  title={A fast solver for the narrow capture and narrow escape problems in the sphere},
  author={Kaye, Jason and Greengard, Leslie},
  journal={Journal of Computational Physics: X},
  volume={5},
  pages={100047},
  year={2020},
  publisher={Elsevier}
}

@article{chen2011,
  title={Asymptotic analysis for the narrow escape problem},
  author={Chen, Xinfu and Friedman, Avner},
  journal={SIAM journal on mathematical analysis},
  volume={43},
  number={6},
  pages={2542--2563},
  year={2011},
  publisher={SIAM}
}

@article{vaccario2015,
  title={First-passage times in $d$-dimensional heterogeneous media},
  author={Vaccario, G and Antoine, C and Talbot, J},
  journal={Phys Rev Lett},
  volume={115},
  number={24},
  pages={240601},
  year={2015},
  publisher={APS}
}

@book{carslaw1959,
  title={Conduction of heat in solids},
  author={Carslaw, Horatio Scott and Jaeger, John Conrad},
  publisher={Oxford: Clarendon Press},
  edition={2},
  year={1959}
}

@article{lawley2019dtmfpt,
title={Diffusive search for diffusing targets with fluctuating diffusivity and gating},
author={Lawley, S D and Miles, C E},
journal={Journal of Nonlinear Science},
year={2019},
  doi = {10.1007/s00332-019-09564-1},
  url = {https://doi.org/10.1007/s00332-019-09564-1},
  note={https://doi.org/10.1007/s00332-019-09564-1}
}

@article{schuss_narrow_2007,
	title = {The narrow escape problem for diffusion in cellular microdomains},
	volume = {104},
	url = {http://www.pnas.org/content/104/41/16098.short},
	number = {41},
	urldate = {2015-08-10},
	journal = {Proceedings of the National Academy of Sciences},
	author = {Schuss, Z. and Singer, A. and Holcman, David},
	year = {2007},
	pages = {16098--16103}
}

@article{lindsay2015,
	title = {Narrow escape problem with a mixed trap and the effect of orientation},
	volume = {91},
	issn = {1539-3755, 1550-2376},
	url = {http://link.aps.org/doi/10.1103/PhysRevE.91.032111},
	doi = {10.1103/PhysRevE.91.032111},
	language = {en},
	number = {3},
	urldate = {2015-08-10},
	journal = {Phys Rev E},
	author = {Lindsay, A. E. and Kolokolnikov, T. and Tzou, J. C.},
	month = mar,
	year = {2015}
}

@article{ward1993,
  title={Strong localized perturbations of eigenvalue problems},
  author={Ward, M J and Keller, J B},
  journal={SIAM J Appl Math},
  volume={53},
  number={3},
  pages={770--798},
  year={1993},
  publisher={SIAM}
}

@book{jackson1975,
	address = {New York},
	edition = {2nd edition},
	title = {Classical {Electrodynamics}},
	isbn = {978-0-471-43132-9},
	language = {English},
	publisher = {Wiley},
	author = {Jackson, J D},
	month = oct,
	year = {1975}
}

@article{gomez2015,
  title={Asymptotic analysis of narrow escape problems in nonspherical three-dimensional domains},
  author={Gomez, D and Cheviakov, A F},
  journal={Phys Rev E},
  volume={91},
  number={1},
  pages={012137},
  year={2015},
  publisher={APS}
}

@article{berez04,
	title = {Boundary homogenization for trapping by patchy surfaces},
	volume = {121},
	doi = {10.1063/1.1814351},
	number = {22},
	journal = {J. Chem. Phys},
	author = {Berezhkovskii, A. M. and Makhnovskii, Y. A. and Monine, M. I. and Zitserman, V. Y. and Shvartsman, S. Y.},
	year = {2004},
	pages = {11390--11394}
}

@article{holcman2014,
	title = {The Narrow Escape Problem},
	volume = {56},
	doi = {10.1137/120898395},
	pages = {213--257},
	number = {2},
	journal = {{SIAM} Rev},
	author = {Holcman, D and Schuss, Z},
	year = {2014}
}

@article{ward10,
	title = {{An asymptotic analysis of the mean first passage time for narrow escape problems: Part I: Two-dimensional domains}},
	volume = {8},
	language = {en},
	number = {3},
	journal = {Multiscale Model Simul.},
	author = {Pillay, S. and Ward, M. J. and Peirce, A. and Kolokolnikov, T.},
	year = {2010},
	pages = {803--835}
}

@article{ward93,
	title = {Summing logarithmic expansions for singularly perturbed eigenvalue problems},
	volume = {53},
	language = {en},
	number = {3},
	journal = {SIAM J. Appl. Math},
	author = {Ward, M. J. and Heshaw, W. D. and Keller, J. B.},
	year = {1993},
	pages = {799--828}
}

@article{PB3,
	title = {Escape from subcellular domains with randomly switching boundaries},
	journal = {Multiscale Model Sim},
	author = {Bressloff, P C and Lawley, S D},
	year = {2015},
	pages = {1420--1445},
	volume = {13},
	number = {4}
}

@article{bressloff2017temporal,
  title={Temporal disorder as a mechanism for spatially heterogeneous diffusion},
  author = {Bressloff, P C and Lawley, S D},
  journal = {Phys Rev E - Rapid Comm},
  volume={95},
  number={6},
  pages={060101},
  year={2017},
  publisher={APS}
}

@article{bernoff_boundary_2018,
    title = {Boundary {Homogenization} and {Capture} {Time} {Distributions} of {Semipermeable} {Membranes} with {Periodic} {Patterns} of {Reactive} {Sites}},
    volume = {16},
    issn = {1540-3459},
    url = {http://epubs.siam.org/doi/abs/10.1137/17M1162512},
    doi = {10.1137/17M1162512},
    number = {3},
    urldate = {2018-09-29},
    journal = {Multiscale Modeling \& Simulation},
    author = {Bernoff, A. and Lindsay, A. and Schmidt, D.},
    month = jan,
    year = {2018},
    pages = {1411--1447},
}

@article{linn_extreme_2022,
    title = {Extreme hitting probabilities for diffusion*},
    volume = {55},
    issn = {1751-8113, 1751-8121},
    url = {https://iopscience.iop.org/article/10.1088/1751-8121/ac8191},
    doi = {10.1088/1751-8121/ac8191},
    language = {en},
    number = {34},
    urldate = {2024-04-22},
    journal = {Journal of Physics A: Mathematical and Theoretical},
    author = {Linn, Samantha and Lawley, Sean D},
    month = aug,
    year = {2022},
    pages = {345002},
}
\bibliographystyle{siam}

\end{document}